\documentclass[aip,apl,superscriptaddress,reprint]{revtex4-1}
\pdfoutput=1
\usepackage[version=3]{mhchem} 
\usepackage[dvips]{epsfig}
\usepackage{graphicx}  
\usepackage{dcolumn}   
\usepackage{bm}        
\usepackage{amsmath}   
\usepackage{amssymb}   
\usepackage{color}
\usepackage{float}
\usepackage{hyperref}
\hypersetup{
     colorlinks   = true,
     citecolor    = blue,
     linkcolor    = red,
     urlcolor     = black
     }
\usepackage{etoolbox}
\patchcmd{\section}
  {\centering}
  {\raggedright}
  {}
  {}
\patchcmd{\subsection}
  {\centering}
  {\raggedright}
  {}
  {}

\begin{document}

\title{Optomechanically amplified wavelength conversion in diamond microcavities}

\author{Matthew Mitchell}
\affiliation{Department of Physics and Astronomy and Institute for Quantum Science and Technology, University of Calgary, Calgary, AB, T2N 1N4, Canada}

\author{David P.  Lake}
\affiliation{Department of Physics and Astronomy and Institute for Quantum Science and Technology, University of Calgary, Calgary, AB, T2N 1N4, Canada}

\author{Paul E. Barclay}
\email{pbarclay@ucalgary.ca}
\affiliation{Department of Physics and Astronomy and Institute for Quantum Science and Technology, University of Calgary, Calgary, AB, T2N 1N4, Canada}

\begin{abstract}
\noindent Efficient, low noise conversion between different colors of light  is a necessary tool for interfacing quantum optical technologies that have different operating wavelengths. Optomechanically mediated wavelength conversion and amplification is a potential method for realizing this technology, and is demonstrated here in microdisks fabricated from single crystal diamond--a material that can host a wide range of quantum emitters. Frequency up--conversion is demonstrated with internal conversion efficiency of $\sim$45\% using both narrow and broadband probe fields, and optomechanical frequency conversion with amplification is demonstrated in the optical regime for the first time.
\end{abstract}

\maketitle

\noindent

The interaction of light and vibration of matter is a rich area of physics dating to Brillouin and Raman's  studies of scattering phenomena \cite{ref:brillouin1922ddl, ref:raman1928anr}. With the advent of quantum mechanics, the concept of  vibrational energy quanta, phonons \cite{ref:tamm1932otp}, spurred the study of photon--phonon scattering processes, leading to many technological breakthroughs in spectroscopy, medicine, and communications. Photon-phonon interactions are becoming increasing relevant for solid state quantum technologies, which utilize phonons for transducing \cite{ref:gustafsson2014ppc, ref:chu2017qas}, storing \cite{ref:lee2011mnc,ref:england2015sar}, and transmitting information \cite{ref:patel2018smp}. Cavity optomechanics aims to enhance coherent phonon-photon interactions through co--localization of mechanical and optical resonances coupled via radiation pressure or other optical forces \cite{ref:aspelmeyer2014co}. These systems provide a platform that could have application in a quantum network or quantum internet \cite{ref:kimble2008qi}, namely as a phonon mediated way of transducing quantum information from visible or microwave photons to telecommunication wavelength photons \cite{ref:andrews2014bae,ref:habraken2012cmc,ref:simon2017tag}. Phonon mediated wavelength conversion has been demonstrated in both optical \cite{ref:hill2012cow, ref:dong2012odm, ref:liu2013eit} and microwave \cite{ref:palomaki2013cst, ref:andrews2014bae} regimes, and many current efforts are focused on optical to microwave transduction \cite{ref:bochmann2013ncb, ref:rueda2016emt, ref:vainsencher2016bdc, ref:reed2017fco, ref:higginbotham2018heo, ref:fan2018sce}.

\begin{figure}[t]
  \centering
  \includegraphics[width=\linewidth]{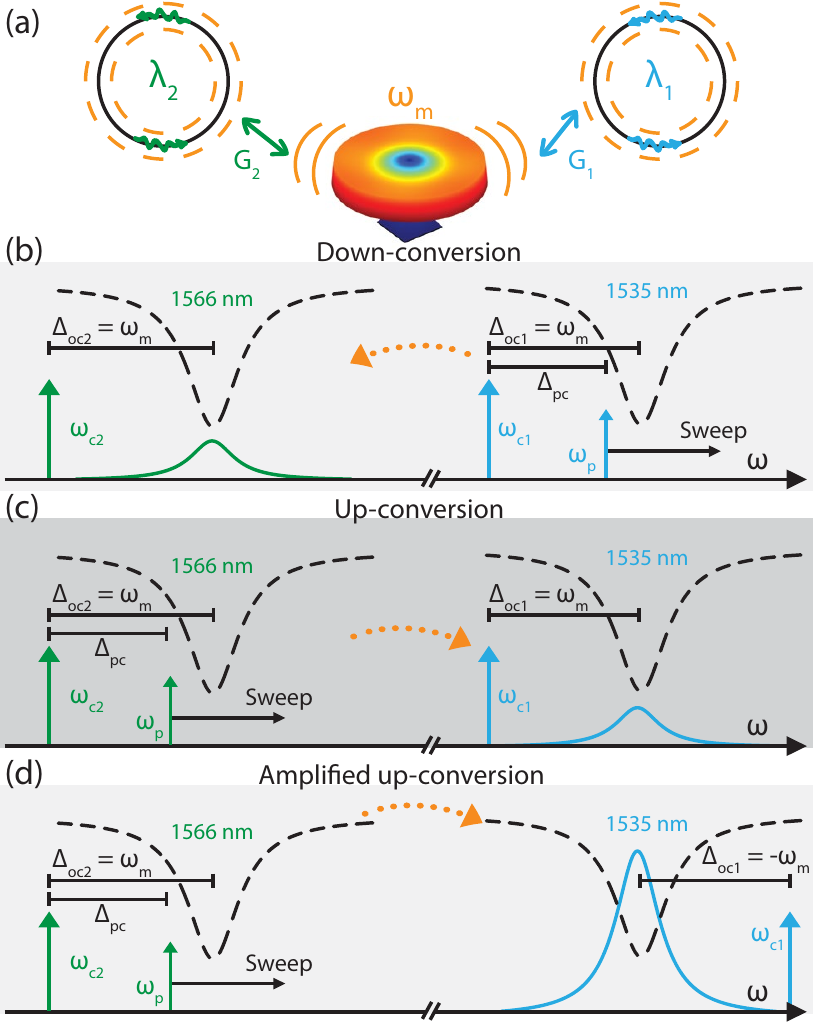}
 \caption{(a) Illustration of the system under study. Two microdisk optical whispering gallery mode resonances at $\lambda_1$ and $\lambda_2$ are coupled to the microdisk's mechanical radial breathing mode whose frequency is $\omega_\text{m}$. Schematics illustrating the cavity density of states, control and probe field detunings for the case of frequency up-- and down--conversion with no amplification (b,c), and (d) frequency up--conversion with amplification.}
 \label{fig:Cartoon}
\end{figure}

Recently, nanofabrication advances have enabled initial investigations of wide electronic bandgap materials such as single--crystal diamond (SCD) for cavity optomechanics \cite{ref:mitchell2016scd, ref:burek2016doc, ref:lake2018oit}. Diamond does not suffer from nonlinear absorption at telecommunications wavelengths, which can limit achievable local field strength and corresponding optomechanical coupling rates in smaller bandgap materials such as silicon \cite{ref:barclay2005nrs, ref:chan2011lcn}. A bonus of working with diamond is that it also hosts highly coherent artificial atoms such as the nitrogen vacancy (NV) and silicon vacancy (SiV) color centers \cite{ref:aharonovich2016ssspe}, which have been used in demonstrations of quantum memory \cite{ref:pfender2017pad,ref:sukachev2017svs}, quantum entanglement \cite{ref:togan2010qeb,ref:bernien2013heb}, quantum teleportation \cite{ref:pfaff2014uqt}, loophole-free Bell's inequality violation \cite{ref:henson2015lfbe}, and the transfer of phase information from microwave to optical fields \cite{ref:lekavicius2017top}. Diamond cavity optomechanics can in principle interface these color centers with light and other quantum systems in new ways \cite{ref:habraken2012cmc,ref:simon2017tag}. One such application is conversion of color center emission to telecommunication wavelengths. A purely optomechanical approach allows  high efficiency wavelength conversion between any two wavelengths of a cavity's mode spectrum \cite{ref:safavi-naeini2011pfa,ref:hill2012cow, ref:dong2012odm, ref:liu2013eit} for relatively low input power, negating the need for material dependent nonlinear optical processes \cite{ref:dreau2018qfc,ref:farrera2016ncb, ref:maring2017pqs,ref:rakher2010qtt,ref:pelc2012dqi,ref:degreve2012qds,ref:radnaev2010aqm,ref:li2016eal}. Here we demonstrate optomechanical wavelength conversion in a diamond cavity for the first time, and show that this scheme can be operated in a regime where the converted signal is optomechanically amplified. We also probe the bandwidth and gain properties of this process using a novel broadband coherent spectroscopy technique.

\begin{figure*}[t]
  \centering
  \includegraphics[width=\linewidth]{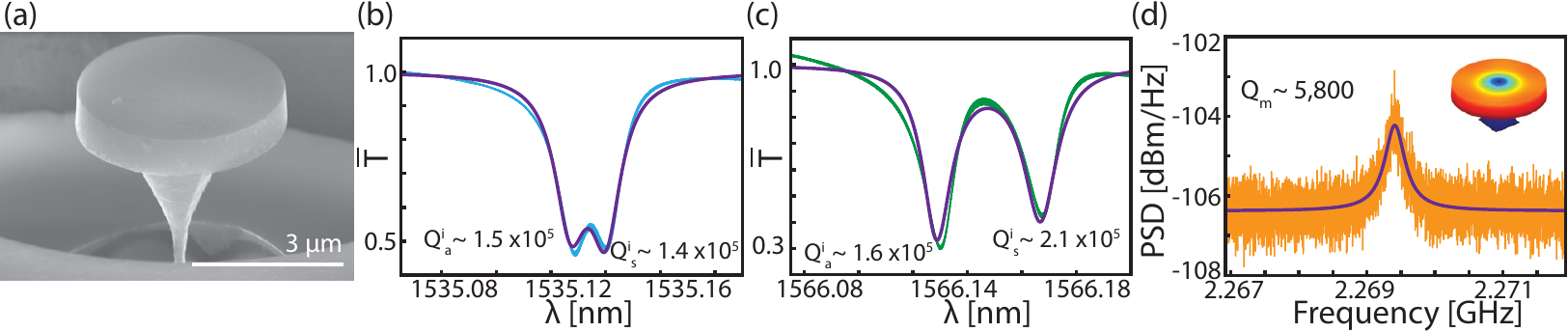}
 \caption{(a) Scanning electron micrograph of a diamond microdisk similar to the device under study here ($\sim 5\ \mu$m diameter). (b,c) Optical whispering gallery mode resonances used in the frequency conversion process. Intrinsic optical quality factors of the symmetric ($Q_\text{s}^{i}$) and anti--symmetric ($Q_\text{a}^{i}$) doublet modes extracted from fit lineshapes are indicated. (d) Photodetected power spectral density of the optomechanically transduced thermally driven microdisk RBM motion in ambient pressure and temperature  (Inset: COMSOL simulated mode profile). Measured at input power sufficiently low to not affect the mechanical resonance dynamics.  Mechanical quality factor $Q_\text{m}\sim 5,800$ is extracted from the fit lineshape.}
 \label{fig:Modes}
\end{figure*}

Cavity optomechanical wavelength conversion coherently couples two optical cavity modes at frequencies $\omega_\text{o,1}$ and $\omega_\text{o,2}$ via their independent optomechanical coupling to a common cavity mechanical resonance at frequency $\omega_\text{m}$. In microdisk cavities, such as the diamond cavity  studied here and illustrated in Fig.\ \ref{fig:Cartoon}(a), the optomechanical coupling is created by the radiation pressure force exerted by optical whispering gallery modes on the microdisk.  This interaction can coherently exchange energy between  optical and mechanical domains when a strong control laser  is input to the cavity at a frequency $\omega_\text{c}$ red-detuned $\Delta_\text{oc} = \omega_\text{o}-\omega_\text{c} = \omega_\text{m}$ from the cavity mode  by the mechanical resonance frequency. The resulting system can be represented by a photon-phonon beamsplitter Hamiltonian \cite{ref:aspelmeyer2014co}, and has been used to demonstrate optomechanically induced transparency (OMIT) and slow light \cite{ref:weis2010oit, ref:safavi2011eit} in a variety of devices, including in diamond microdisks \cite{ref:lake2018oit}.

In multimode cavity optomechanical systems, coherent photon-phonon coupling is harnessed to convert probe photons input on resonance with a mode at  $\omega_\text{o,1}$ to a mechanical excitation that is in turn converted to signal photons resonant with the second cavity mode $\omega_\text{o,2}$. This process requires inputting two control lasers, referred to as write and read fields, to the cavity. Typically both control lasers are red--detuned by $\Delta_\text{oc} = \omega_\text{m}$ from their respective cavity modes  as illustrated in Figs.\ \ref{fig:Cartoon}(b,c).  This setup, which has been used in previous optical domain wavelength conversion demonstrations \cite{ref:hill2012cow, ref:dong2012odm, ref:liu2013eit},  can in principle be amplified by blue-detuning the read laser from the cavity ($\Delta_\text{oc} = -\omega_\text{m}$), as shown in Fig.\ \ref{fig:Cartoon}(d). This enables low--noise amplification of the converted photons, as recently demonstrated by Ockeloen-Korppi et al.\ in the microwave regime \cite{ref:ockeloen-korppi2016lna} and reported below for the first time in the optical domain. Such two--port amplification  avoids optomechanical self oscillation by balancing optomechanical amplification with damping for the blue-- and red--detuned optical modes, respectively, and unlike traditional optomechanical amplification \cite{ref:massel2011man, ref:mcrae2012nta, ref:li2012mco} has a fundamentally unlimited gain--bandwidth product.

\begin{figure}
  \centering
  \includegraphics[width=\linewidth]{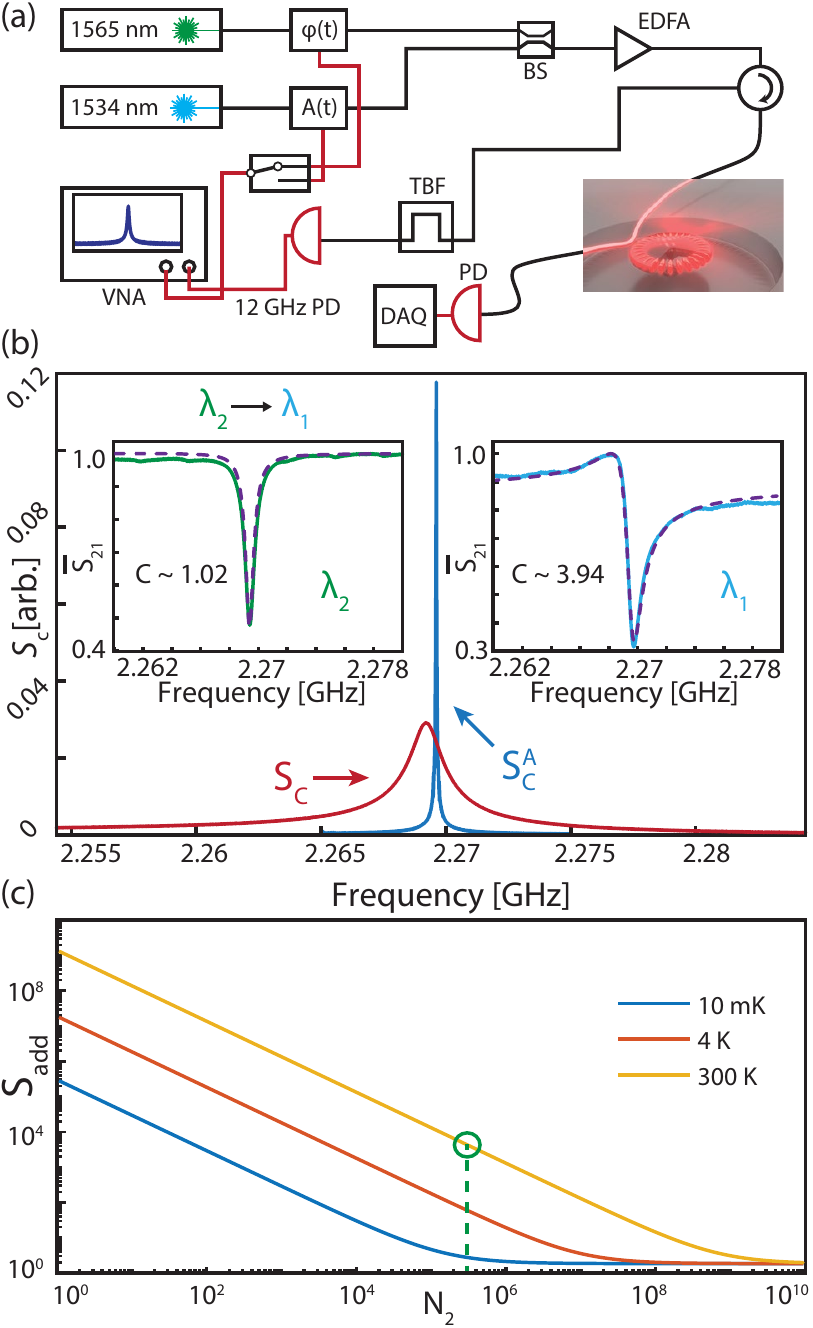}
 \caption{(a) Experimental setup used for wavelength conversion and amplification. Phase and amplitude EOM's driven by the vector network analyzer (VNA) are used to generate the probe fields from the control fields where an RF switch controls which laser to modulate. A 50\%/50\% waveguide coupler combines the input fields which are coupled to the microdisk via a dimpled tapered fiber. A tunable band pass filter (TBF) is used to filter the output of the cavity and the photodetected signal is analyzed by the VNA. (b) Beat note between converted photons and control field of the same color measured on the VNA for frequency up--conversion with and without amplification. OMIT spectra for the $\lambda_2$ and $\lambda_1$ optical modes are shown as insets, where the cooperativity is extracted from the depth of the OMIT feature. (c) Predicted added noise to amplified signal based on Eq.\ \ref{eqn:noise}, and system parameters, with operating regime circled.}
 \label{fig:OMIT}
\end{figure}

The diamond microdisks used here for wavelength conversion, an example of which is shown in the scanning electron micrograph image in Fig.\ \ref{fig:Modes}(a), support whispering gallery modes with optical quality factor $Q_\text{o}$ sufficiently high for operation in the resolved sideband regime of cavity optomechanics.  These devices were fabricated using an optimized reactive ion undercutting process described in Ref.\ \cite{ref:mitchell2018rqo} that results in a smooth diamond microdisk supported by a thin diamond pillar, and have higher $Q_\text{o}$ than those previously used for single-mode diamond optomechanics studies \cite{ref:lake2018oit}.

All of the measurements presented below involve two sets of microdisk optical modes widely separated in wavelength and mutually coupled to the same mechanical radial breathing mode (RBM) of the microdisk, whose mechanical displacement profile is illustrated in Fig.\ \ref{fig:Cartoon}(a). Their optomechanical properties were probed using coherent fiber taper optical mode spectroscopy as in previous work \cite{ref:mitchell2016scd, ref:mitchell2018rqo, ref:lake2018oit} using the apparatus described in more detail below.  Fiber taper transmission spectra for the two modes are shown in Figs.\ \ref{fig:Modes}(b,c), revealing resonances with frequencies $\omega_\text{o,1}/2\pi = 195$ THz ($\lambda_1 = 1535$ nm) and $\omega_\text{o,2}/2\pi = 191$ THz ($\lambda_2 = 1566$ nm), and unloaded $Q_\text{o}^\text{i} \sim 1.4\times 10^5$ and $2.1\times 10^5$, respectively.  The doublet nature of the resonances indicates that they are standing wave modes  \cite{ref:borselli2005brs}; the  red mode of each doublet was used for all of the measurements described below.  The $\lambda_1$ mode was found to be TM--like, while the $\lambda_2$ mode was TE--like, which is reflected in the different splitting of each doublet, but is not an essential ingredient for the results presented here.  These modes are used to optomechanically transduce the motion of the microdisk mechanical resonances, with a typical photodetected power spectral density of the thermomechanical motion of the microdisk's RBM shown in Fig.\ \ref{fig:Modes}(d), revealing a resonance frequency $\omega_\text{m}/2\pi = 2.27$ GHz and mechanical quality factor $Q_\text{m} = \omega_\text{m}/\Gamma_\text{m} \sim 5,800$ in the ambient conditions used for all of the reported measurements. The device's combination of high mechanical frequency and high $Q_\text{o}$ places it in the resolved sideband regime $\kappa/\omega_\text{m} \leq$ 1, where $\kappa = \omega_\text{o}/Q_\text{o}$ is the total cavity photon decay rate.

The experimental setup used to coherently couple light at multiple wavelengths to the microdisk's mechanical motion is shown in Fig.\ \ref{fig:OMIT}(a).  Two independent tunable diode lasers (Newport TLB-6700) generate the control fields. An amplitude  or phase electro--optic modulator (EOSpace) driven by a vector network analyzer (Keysight E5063A) at a swept frequency $\delta\omega_\text{p}$ generates the probe field by adding sidebands to the $\omega_\text{c,1}$ or $\omega_\text{c,2}$  control field, respectively. An RF switch selects which control laser will be modulated, determining whether to carry out frequency down- or up--conversion, respectively. The control field intensities are amplified via an erbium--doped fiber amplifier (EDFA, Pritel) before being coupled to the microdisk via the optical fiber taper. The transmitted signal is used for the optical mode spectroscopy described above, while the reflected signal is routed via an optical circulator through a tunable band pass filter (TBF, Optoplex, 100 GHz) that separates the converted signal and the read control field from the write control and the input probe fields. The beat note from the converted signal field interfering with the read control field is then demodulated from the photodetector output by the vector network analyzer.

To demonstrate wavelength conversion the lasers were first setup in the standard up-conversion configuration illustrated in Fig.\ \ref{fig:Cartoon}(c): both read and write control fields red--detuned from their respective cavity modes by $\Delta_\text{oc,1} = \Delta_\text{oc,2} = \omega_\text{m}$, and the write control field weakly modulated at $\delta\omega_\text{p}$ to generate the probe sideband  (wavelength $\lambda_2$). The resulting measured beatnote, $S_\text{C}(\delta\omega_\text{p})$, between the up-converted field at $\lambda_1$ and the read control field,  is shown in Fig.\ \ref{fig:OMIT}(b). Efficient conversion is observed when $\delta\omega_\text{p} = \omega_\text{m}$, i.e.\ when the probe field is on-resonance with the cavity mode, and is observed over a bandwidth defined by the optomechanically broadened mechanical linewidth.

Calibrating $S_\text{C}$ and writing it as a wavelength conversion efficiency requires careful characterization of the detectors and electronics used in the experiment. Instead,  following previous reports \cite{ref:hill2012cow, ref:liu2013eit}, we infer an external conversion efficiency $\eta_\text{ext} = \eta_1\eta_2\eta_\text{int}$ between $\lambda_2$ and $\lambda_1$ from the optomechanical properties of our device. Here $\eta_1=13\%$ and $\eta_2=21\%$ are the experimentally measured waveguide-cavity coupling efficiencies at wavelengths $\lambda_1$ and $\lambda_2$ respectively, extracted by determining the coupling rate $\kappa_\text{e}$ between the microdisk standing wave modes and  the forward propagating fiber mode from fits to the doublet resonances, taking into account the standing wave modes' equal coupling to the backwards propagating fiber mode. The internal conversion efficiency, $\eta_\text{int} = 4C_1 C_2/(1+C_1+C_2)^2$, is determined solely by the optomechanical cooperativity, $C_j$, of each mode $\lambda_j$ \cite{ref:hill2012cow}. Here $C_j=4N_j g_{0,j}^2/\kappa_j\Gamma_\text{m}$ was measured from the modes' OMIT dips shown in the insets to Fig.\ \ref{fig:OMIT}(b), where $N_j$ and $g_{0,j}$ are the respective control field intracavity photon number and the vacuum optomechanical coupling rate of mode $j$. These single color OMIT measurements record the beatnote between control field $j$ in the red detuned configuration ($\Delta \omega_{\text{oc},j} = \omega_\text{m}$) and its modulator generated sideband at $\omega_{\text{c},j} + \delta\omega_\text{p}$. Note that while both control lasers were kept on during this measurement, only one OMIT process was carried out at a time, by placing the other laser off resonance. By fitting the OMIT spectra in Fig.\ \ref{fig:OMIT}(b) to the model described in Supplement 1 we extract $C_1 \sim 3.94$ and $C_2 \sim 1.02$, for $N_1\sim 3.1\times10^6$ and $N_2\sim 2.2\times10^5$, respectively, corresponding to $\eta_\text{int} \sim 45\%$. Here $N_j$ and $\Delta_{\text{oc},j}$ were used as the only fitting parameters where all other required parameters were measured independently. Measurements of $g_0$ were performed using the phase tone calibration method \cite{ref:gorodetksy2010dvo}, giving $g_0/2\pi \sim$ 15 kHz and 27 kHz for the $\lambda_1$ and $\lambda_2$ modes, respectively, as described in Supplement 1. Note that the Fano--shape of the $\lambda_1$ OMIT dip is related to a non-zero phase imparted by the amplitude modulator, and the demodulation process of the VNA, as described in detail in Supplement 1. These measured values for $C_j$ are consistent with the estimated fiber taper input power for each control field of $P_\text{in} \sim$ 16 mW and $\sim$ 3.8 mW.

Increasing  $\eta_\text{int}$ could be achieved by operating in the $C_j \gg 1$ regime and by matching $C_1 = C_2$. For the diamond cavities studied here, in practice $N_j$ can be increased above $10^6$ until becoming limited by thermal instabilities within the cavity. For example, C > 3.9 is demonstrated for measurements presented in Supplement 1. In this experiment we were unable to balance $C_1$ and $C_2$ while maintaining $C_j > 1$ as the input laser used to drive $\lambda_2$ mode was less efficiently amplified in the EDFA. Combined with greater loss in the system at $\lambda_2$, this resulted in $N_2 < N_1$.  The use of a second EDFA with a gain maximum near $\lambda_2$ would allow independent control of $N$ for each mode. Assuming each mode could reach $C\sim2$,  $\eta_\text{int}\sim 64\%$ would be achievable.  Finally, the relatively small $\eta_\text{ext}=1.23\%$ demonstrated here could be improved by operating in the over coupled regime, which requires either higher $Q_\text{o}$ or improved coupling via, for example, an external on-chip waveguide.  Unfortunately the $\omega_\text{o,2}$ mode was outside the range of the TBF, prohibiting measurement of down--conversion for this device. However, down--conversion was measured using a different device, with lower conversion efficiency, as described in Supplement 1.

These devices are also promising for quantum limited amplification of the wavelength converted signal field, as previously demonstrated in the microwave domain \cite{ref:ockeloen-korppi2016lna}.  As described above and shown in Fig.\ \ref{fig:Cartoon}(d), signal amplification is achieved by placing the $\lambda_1$ read control laser blue--detuned from the cavity, $\Delta_\text{oc,1} = -\omega_\text{m}$, while keeping the $\lambda_2$ write control laser red--detuned as before, $\Delta_\text{oc,2} = \omega_\text{m}$. This results in what can be thought of as a two step process involving a beam--splitter interaction between the $\lambda_2$ mode and the mechanical mode, and parametric amplification between the mechanical mode and the $\lambda_1$ mode  \cite{ref:ockeloen-korppi2016lna}.
The converted--amplified signal is detected in the same fashion as above, where the beat note $S_\text{C}^\text{A}$ produced by the amplified converted light and the read control field is shown in Fig.\ \ref{fig:OMIT}(b). The frequency and linewidth of the amplified, $S_\text{C}^\text{A}$, and unamplified, $S_\text{C}$, spectra is governed by the optomechanical damping ($\Gamma_\text{opt}$) and optical spring effect ($\delta\omega_\text{m}$) induced on the mechanical resonance dynamics by each control laser. As both control lasers are red--detuned for $S_\text{C}$, its effective mechanical linewidth is broader and the center frequency is red--shifted compared to its intrinsic value shown in Fig.\ \ref{fig:Modes}(d). In contrast, as described above, for $S_\text{C}^\text{A}$ the $\lambda_1$ laser is blue--detuned while $\lambda_2 $ is red--detuned. As the photon assisted optomechanical coupling rates, $G_j = g_{0,j}\sqrt{N_j}$, are such that $G_1 > G_2$ ($G_1 = 2\pi \times 26.4\ \text{MHz} , G_2 = 2\pi \times 12.5\ \text{MHz} $), this initially resulted in the microdisk being in the self-oscillation regime ($\Gamma_\text{m}+\Gamma_\text{opt} < 0$). To obtain the amplified spectra in Fig.\ \ref{fig:OMIT}(b) $\Delta_\text{oc,1}$ was adjusted such that the device was no longer in the self--oscillation regime, although a narrowed linewidth and blue--shift of the center frequency is still observed due to the imbalance of $G_1$ and $G_2$. Balancing $G_1$ and $G_2$ was a technical limitation in our experiment, which could be remedied by using a second EDFA to independently tune $G_1$ and $G_2$, enabling operation in the large gain--bandwidth product regime. Ideally, in the low--noise amplification scheme described in Ref.\ \cite{ref:ockeloen-korppi2016lna} the system should be operated in the $G_2 \gtrsim G_1$ regime to avoid instabilities associated with optomechanical self--oscillation.

To characterize the noise properties of the $S_\text{C}^\text{A}$ spectrum we follow the analysis in Ref. \cite{ref:ockeloen-korppi2016lna}, starting with expected frequency--converted gain, given by

\begin{equation}\label{eqn:A_x}
A_\text{x} = 2 \left(\frac{\kappa_\text{e}}{\kappa}\frac{4G_1G_2/\kappa}{\Gamma_\text{m}-4\overline{G}^2/\kappa}\right),
\end{equation}

\noindent where the optical cavity dissipation constants have been taken to be equal and $\overline{G}^2 = G_1^2 - G_2^2$. The adjusted $G_1$ value was determined from the linewidth of $S_\text{C}^\text{A}$ where $G_2$ was held constant, giving $G_1 \sim 2\pi \times 14.1$ MHz. For the operating conditions here, and taking $\kappa = \kappa_2$, Eqn.\ \ref{eqn:A_x} gives $A_\text{x} \sim 1.2$. For these operating conditions the amplifier is not in a low--noise regime as the added noise to $S_\text{C}^\text{A}$ is dominated by the contribution from the mechanical bath, whose equivalent added noise on resonance is given by $S_\text{add,m}/|A_\text{x}|^2$:

\begin{equation}\label{eqn:noise_mech}
S_\text{add,m} = \frac{\Gamma_\text{m}\kappa_\text{e}G_1^2}{\left|\Gamma_\text{m}\kappa/4 - \overline{G}^2\right|^2}\left(n_\text{th}+\frac{1}{2}\right),
\end{equation}

\noindent where $n_\text{th}\sim k_\text{B}T/\hbar\omega_\text{m}$ is the thermal occupation of the RBM. For this experiment this results in $\sim 6.52 \times 10^3$ added quanta to the $S_\text{C}^\text{A}$ spectra. To predict future performance of this system within the context of low--noise single photon wavelength conversion, we calculate the added noise to the amplified optical signal in the ideal high gain ($G_2 \gtrsim G_1$) and large $\Gamma_\text{opt}$ regime, given by

\begin{equation}\label{eqn:noise}
S_\text{add} = \frac{\Gamma_\text{m}\kappa}{4G_2^2}\frac{\kappa}{\kappa_\text{e}}\left(n_\text{th}+\frac{1}{2}\right)+\frac{\kappa_\text{i}}{\kappa_\text{e}} + \frac{1}{2}.
\end{equation}

\noindent Here $\kappa_\text{i} = \kappa - 2\kappa_\text{e}$ is the intrinsic cavity photon decay rate, where the optical decay rates have been taken to be similar for each cavity. The predicted added noise as a function of $N_2$ is shown in Fig.\ \ref{fig:OMIT}(c) for the parameters of the device under study and for various temperatures. We predict that the minimum added noise of $\sim$ 2 quanta, limited by the coupling efficiency achieved here, can be approached for $N_2\sim1\times 10^6$ at 10 mK and $N_2\sim1\times 10^8$ at 4 K for this device, provided that operation in the ($G_2 \gtrsim G_1$) regime is enabled by overcoming technical limitations in our experiment. Furthermore, amplification at the SQL of 1/2 quanta could be possible by improving the coupling to the device such that $\kappa_\text{e} \gg \kappa_\text{i}$.

\begin{figure}[t]
  \centering
  \includegraphics[width=\linewidth]{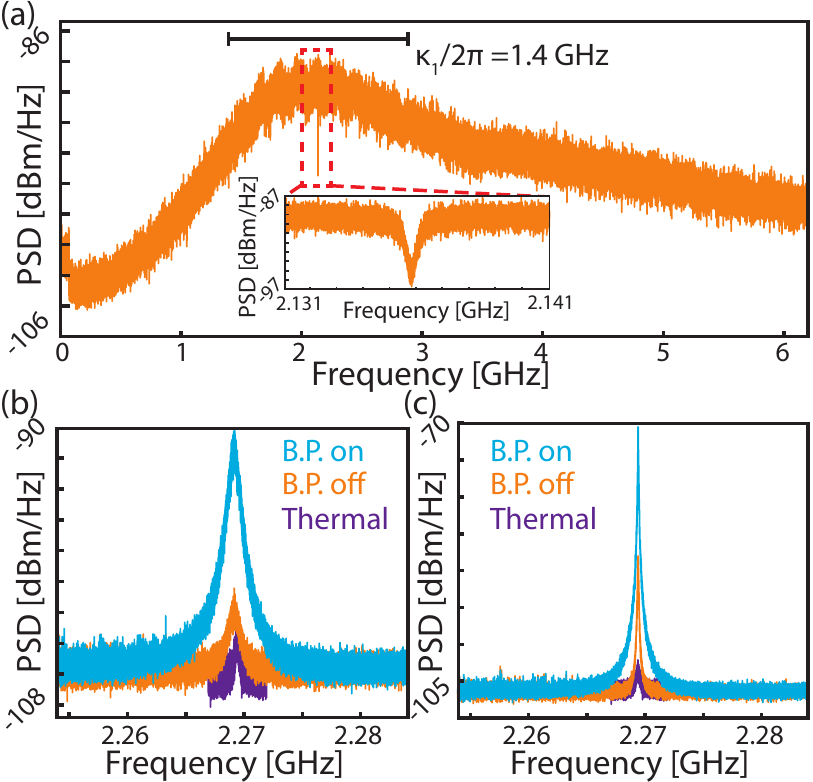}
 \caption{Measurement of OMIT and wavelength conversion via the RSA. (a) Broadband and narrowband (inset) spectrum showing the cavity response and the OMIT feature at $\omega_\text{m}/2\pi$ when a phase EOM is driven with a broadband RF noise source for the device shown in Supplement 1. (b,c) Wavelength conversion for the same device in Fig.\ \ref{fig:OMIT} without (b) and with (c) amplification as measured on the RSA, with the noisy broadband probe (B.P.) field. The thermal motion of the RBM is also shown, which was measured at low power with a single laser to avoid optomechanical back action. Here the noise floor is set by the photodetector whose noise equivalent power, NEP $= 33 \text{pW}/\sqrt{\text{Hz}}$.}
 \label{fig:Noise}
\end{figure}

To further characterize the converted amplified signal we used a broadband (compared to $\kappa$) RF noise source to drive the phase modulator generating the OMIT probe field, similar to the approach described in Ref.\ \cite{ref:zou2017qhe}. Here an arbitrary waveform generator (AWG:Tektronix 70002A) outputting a pseudo Gaussian white noise signal with a sampling rate and length of 25 GS/s and 2 GS, respectively,  drove the RF input of the phase EOM (20 GHz bandwidth) to create a broadband optical probe field from the write field control laser. When the control laser is tuned to $\Delta\omega_\text{oc,2} = \omega_\text{m}$,  observation of both OMIT and wavelength conversion on a real time spectrum analyzer (Tektronix RSA5106B) is possible. Figure\ \ref{fig:Noise}(a) shows the measured probe field's power spectral density, revealing the cavity response for frequency components of the probe that do not coherently drive the mechanics, which in these measurements is the broad peak in the demodulator signal with a line width of $\kappa/2\pi = 1.4$ GHz, and an OMIT dip at  $\delta\omega_\text{p} = \omega_\text{m}$, i.e.\ where the OMIT condition is satisfied. This demonstrates a straightforward method for measuring  OMIT and the cavity response simultaneously in frequency space.

Wavelength conversion of the $\lambda_2$ broadband probe generated by the RF noise source, in both the unamplified and amplified configurations, measured in reflection, is shown in Figs.\ \ref{fig:Noise}(b) and \ref{fig:Noise}(c) respectively. The plots in Figs.\ \ref{fig:Noise}(b) and \ref{fig:Noise}(c) show the detected power spectral density of the converted signal at $\lambda_1$ (blue trace) compared for reference with the corresponding measurement when the $\lambda_2$ probe is off (orange trace) so that the detected spectrum is derived entirely from the optomechanically modified motion of the microdisk due to the $\lambda_1$ and $\lambda_2$ control fields. As above, when both control lasers are red--detuned in Fig.\ \ref{fig:Noise}(b) , the effective mechanical linewidth is broader and the center frequency is red--shifted compared to its intrinsic value (purple trace), which was measured at low power with a single optical mode to avoid optomechanical back action. In the amplified conversion case shown in  Fig.\ \ref{fig:Noise}(c) the linewidth is slightly narrowed and shifted to higher frequency due to the fact that $G_1 > G_2$.  In principle this technique could be utilized in balancing $G_j$ by adjusting $N_j$ such that the linewidth of the observed spectra with the broadband probe off is equal to the intrinsic linewidth observed in the absence of the strong control fields. Here the ratio of the amplified to unamplified peak height is $\sim2.4\times$ larger than what was observed for the coherently driven case in Fig.\ \ref{fig:OMIT}(b). While the control field input power of each mode was held constant between these measurements, $\Delta_\text{oc,1}$ and $\Delta_\text{oc,2}$  were manually set for each case, which could account for this discrepancy.

The converted signal power in both amplified and unamplified cases can be computed directly from the area under the spectrum.  When the probe is off we see the added noise from the thermally driven signal level that would have to be overcome by increasing $C_j$ to allow operation at the single--photon level. We estimate the gain utilizing measurement of the cavity transmission profile of the $\lambda_2$ mode on the RSA utilizing the broadband probe and the data in  Figs.\ \ref{fig:Noise}(c) as outlined in the Supplementary Material. Using this method we estimate a gain of $A_\text{x, exp} \leq 3.2$, which combined with the total linewidth, $\Gamma_\text{tot} = \Gamma_\text{m}+\Gamma_\text{opt} = 2\pi\ \times 5.8$ kHz give a gain--bandwidth product, GBW $\leq$ 19 kHz. Possible explanations for the mismatch of $A_\text{x}$ and $A_\text{x, exp}$ include the approximation of similar $\kappa$ and $\kappa_\text{e}$ for each optical mode in calculating $A_\text{x}$, uncertainty in the measured optical output power from the cavity, and variation in $\Delta_\text{oc,1}$ and $\Delta_\text{oc,2}$, as mentioned above. With the data available a measurement of the added noise to the converted signal was not possible, however the thermal component is expected to dominate, as discussed above. To compare the amplified vs. unamplified wavelength conversion, where we have defined the signal ratio, as the difference in integrated power with the broadband probe field on and off, to isolate the converted signal from the thermal component, which is then divided by the off--resonant noise floor. We find that the amplified conversion signal is 1.62 $\times$ larger than without amplification, for the same input broadband probe. With calibration of the probe spectrum, this broadband probe technique could also be used to determine the conversion efficiency, similar to what has been suggested by Liu et al.\ in Ref.\ \cite{ref:liu2013eit} where a coherent drive and RSA were used to characterize the converted signal.

In summary, we have demonstrated a diamond cavity optomechanical system capable of phonon--mediated wavelength conversion with an internal conversion efficiency of $\sim45\%$. While we were limited by an imbalance of $G_1$ and $G_2$, and $\kappa_\text{e} < \kappa$ in this experiment, operation in the stable, high-gain regime $G_2 \gtrsim G_1$ should be possible by utilizing a second EDFA to allow separate control of $G_1$ and $G_2$, and improving waveguide--cavity coupling would enable $A_\text{x} \gg 1$. To the best of our knowledge, this is also the first demonstration of optomechanical amplification, albeit not operating in the low--noise regime, of the converted light in the optical domain. In this configuration this system has promise for reaching a minimum added noise of $\sim2$ quanta during the amplification process by operating at larger photon number and with  $G_2 \gtrsim G_1$. This device can in principle operate at the standard quantum limit (SQL) where a minimum of half an energy quanta of noise is added to the signal by improving the waveguide--cavity coupling efficiency, and has immediate practical use in increasing signal to noise when measuring quantum optomechanical effects \cite{ref:caves1982qln}.  In addition, use of a broadband noisy optical probe field was introduced for both OMIT and wavelength conversion. Finally, as previous studies in these structures \cite{ref:mitchell2016scd} have demonstrated the existence of high-$Q_\text{o}$ optical modes at $\sim 738$ nm and $\sim 637$ nm, near the zero phonon lines of SiV and NV color centers respectively, these devices have potential as a converter of diamond color center emission to telecommunications wavelengths for application in quantum networks.

\section{Funding}
\noindent This work was supported by National Research Council Canada (NRC), Alberta Innovates, National Sciences and Engineering Research Council of Canada (NSERC), and Canada Foundation for Innovation (CFI).

\vspace{5 mm}
\noindent See Supplement 1 for supporting content.

\section*{References}

%

\clearpage

\setcounter{equation}{0}
\setcounter{figure}{0}
\setcounter{section}{0}
\setcounter{subsection}{0}
\setcounter{table}{0}
\setcounter{page}{1}
\makeatletter
\renewcommand{\theequation}{S\arabic{equation}}
\renewcommand{\thetable}{S\arabic{table}}
\renewcommand{\thefigure}{S\arabic{figure}}
\renewcommand{ \citenumfont}[1]{S#1}
\renewcommand{\bibnumfmt}[1]{[S#1]}

\section*{Supplementary Information}

\renewcommand{\theequation}{S\arabic{equation}}

\section{Electro--optic modulation model}

When modelling the predicted output spectra of the optomechanical cavities studied in this work the details of how the probe fields are generated are vital, due to the sensitivity of the cavity to both phase and amplitude fluctuations. A variety of constructions exist for implementing an optical phase or amplitude modulator. The LiNbO$_3$ electro--optic modulators (EOM) utilized in this work were purchased from EOSpace where the amplitude modulator used is Z--cut (pre--chirp) with an alpha chirp parameter of $\alpha_\text{chirp} \sim 0.6 - 0.8$. In practice this results in both amplitude and phase modulation which must be taken into account when considering the optical transmission and reflection by the optical cavity.

A phasor picture can prove useful when visualizing the differences between pure phase and amplitude modulation as illustrated in the attached animations (Supplement 2 \& 3), where inspiration was taken from Ref.\ \cite{supp:ref:bond2017itf}. Here a frame rotating with the carrier frequency is used such that the electric field vector appears stationary. We observe from this animation that for pure phase modulation the generated sidebands are out of phase when re--phasing in the real plane leading to no amplitude modulation, while for the pure amplitude modulation case they are $\pi$ out of phase when re--phasing in the real plane. In practice our amplitude modulator does not provide pure amplitude modulation. To account for this, we allow the phase of the carrier to be a variable dependent on the type of EOM used to develop a model capable of describing an EOM operating between the pure amplitude and phase modulation regime. First we describe optical field amplitude generated by the EOM and then input to the cavity $\alpha_\text{in}$ as:

\begin{equation}
\alpha_\text{in} = \alpha\left(e^{i\phi}+\frac{\beta}{2} e^{i\omega t}+ \frac{\beta}{2} e^{-i\omega t}\right),
\end{equation}

\noindent where $\alpha$ is the carrier optical field amplitude, $\beta$ is the modulation index, and $\omega$ is the modulation frequency. When $\phi = 0 \pm n\pi$ the field for a pure amplitude modulator is described and for $\phi = \pi/2 \pm n\pi$ a pure phase modulator. To use this model to predict the observed cavity output spectra in this work we begin by writing the input optical field frequency components as $\alpha^0_\text{in} = \alpha e^{i\phi}$, $\alpha^+_\text{in} = \frac{\alpha\beta}{2} e^{-i\omega t}$, and $\alpha^-_\text{in} = \frac{\alpha\beta}{2} e^{i\omega t}$. Then the cavity transmission and reflection amplitudes are given by:

\begin{figure*}[ht]
  \centering
  \includegraphics[width=\linewidth]{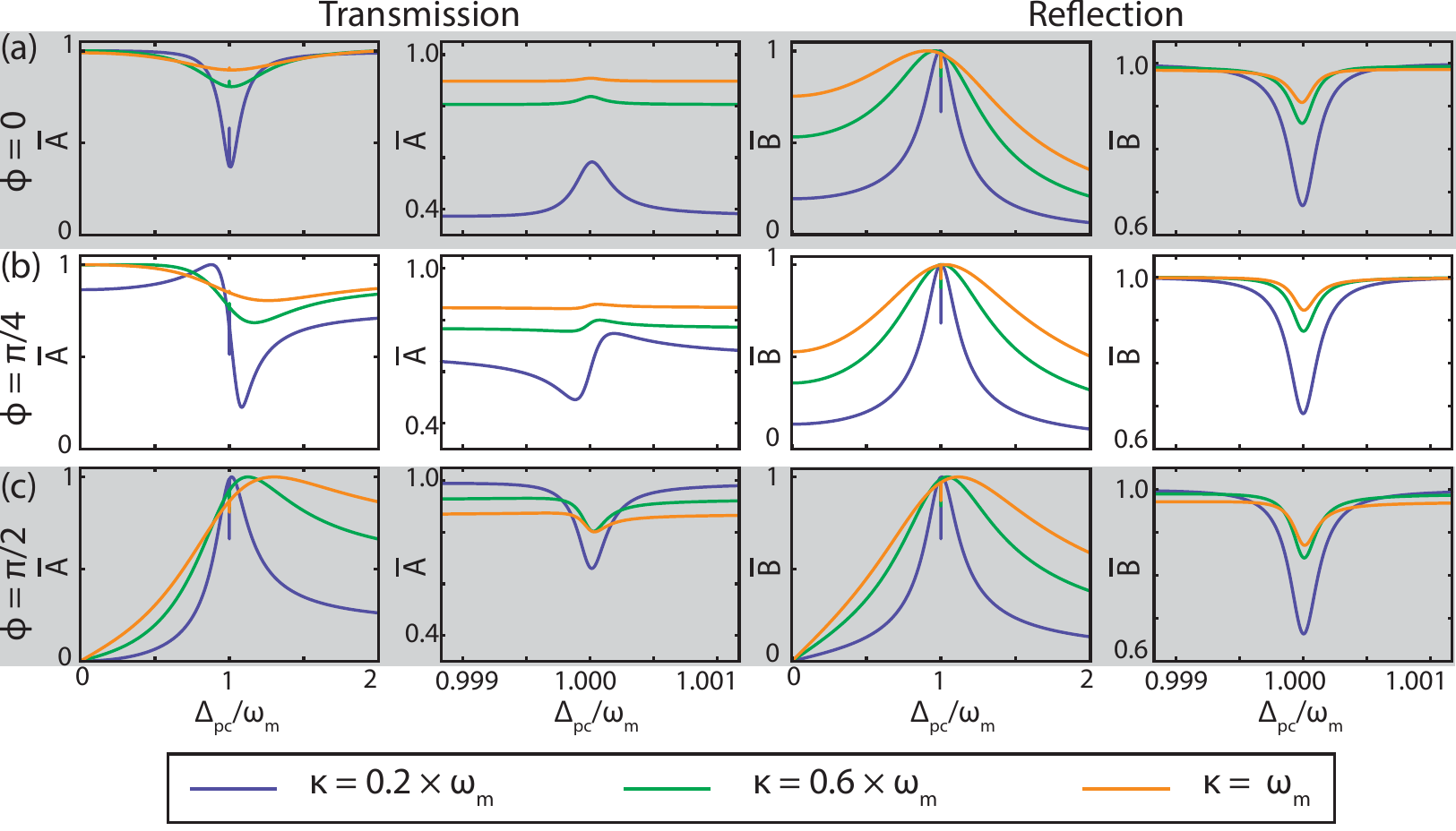}
 \caption{Model for reflection and transmission spectra as a function of probe field detuning, $\Delta_\text{pc}$ and sideband phase parameter $\phi$ for a red--detuned control field with $\Delta_\text{oc} = \omega_\text{m}$, $Q_\text{m} = 5000$, $G = 4.78$ MHz, $\kappa_\text{e} = 0.25\cdot\kappa$, and varying sideband resolution shown in the legend.}
 \label{fig:EOM}
\end{figure*}

\begin{equation}
t_\text{out} = t^+_\text{out} + t^0_\text{out} + t^-_\text{out}
\end{equation}

\noindent and

\begin{equation}
r_\text{out} = r^+_\text{out} + r^0_\text{out} + r^-_\text{out}
\end{equation}

\noindent where

\begin{align}
t^+_\text{out} &= t^+\alpha^+_\text{in} \\ \nonumber
t^0_\text{out} &= t^0\alpha^0_\text{in} \\ \nonumber
t^-_\text{out} &= t^-\alpha^-_\text{in}
\end{align}

\noindent and

\begin{align}
r^+_\text{out} &= r^+\alpha^+_\text{in} \\ \nonumber
r^0_\text{out} &= r^0\alpha^0_\text{in} \\ \nonumber
r^-_\text{out} &= r^-\alpha^-_\text{in}.
\end{align}

\noindent Here the terms ($t^+,t^0,t^-$) and ($r^+,r^0,r^-$) describe the cavity transmission or reflection amplitude, respectively, as seen by each frequency component of the input field, defined as follows. We now consider the power measured by the photodetector in transmission and reflection, $\left|t_\text{out}\right|^2 = t^*_\text{out}t_\text{out}$, and $\left|r_\text{out}\right|^2 = r^*_\text{out}r_\text{out}$:

\begin{equation}
\left|t_\text{out}\right|^2 = |t^0|^2|\alpha_\text{in}^0|^2+t^{0*}t^+\alpha^{0*}_\text{in}\alpha^+_\text{in}+t^0t^{-*}\alpha^{0}_\text{in}\alpha^{-*}_\text{in}+c.c
\end{equation}

\noindent and

\begin{equation}
\left|r_\text{out}\right|^2 = |r^0|^2|\alpha_\text{in}^0|^2+r^{0*}r^+\alpha^{0*}_\text{in}\alpha^+_\text{in}+r^0r^{-*}\alpha^{0}_\text{in}\alpha^{-*}_\text{in}+c.c,
\end{equation}

\noindent where we have neglected terms that are of $\mathcal{O}(\beta^2)$. Substituting our expressions for the input fields gives:

\begin{multline}
\left|t_\text{out}\right|^2 = |t^0|^2\alpha^2 + \alpha^2\frac{\beta}{2}[t^{0*}t^+e^{-i\phi} + t^0t^{-*}e^{i\phi}]e^{-i\omega t} \\
+ \alpha^2\frac{\beta}{2}[t^{0}t^{+*}e^{i\phi} + t^{0*}t^{-}e^{-i\phi}]e^{i\omega t}
\end{multline}

\noindent and

\begin{multline}
\left|r_\text{out}\right|^2 = |r^0|^2\alpha^2 + \alpha^2\frac{\beta}{2}[r^{0*}r^+e^{-i\phi} + r^0r^{-*}e^{i\phi}]e^{-i\omega t} \\
+ \alpha^2\frac{\beta}{2}[r^{0}r^{+*}e^{i\phi} + r^{0*}r^{-}e^{-i\phi}]e^{i\omega t}.
\end{multline}

\begin{figure*}[ht]
  \centering
  \includegraphics[width=\linewidth]{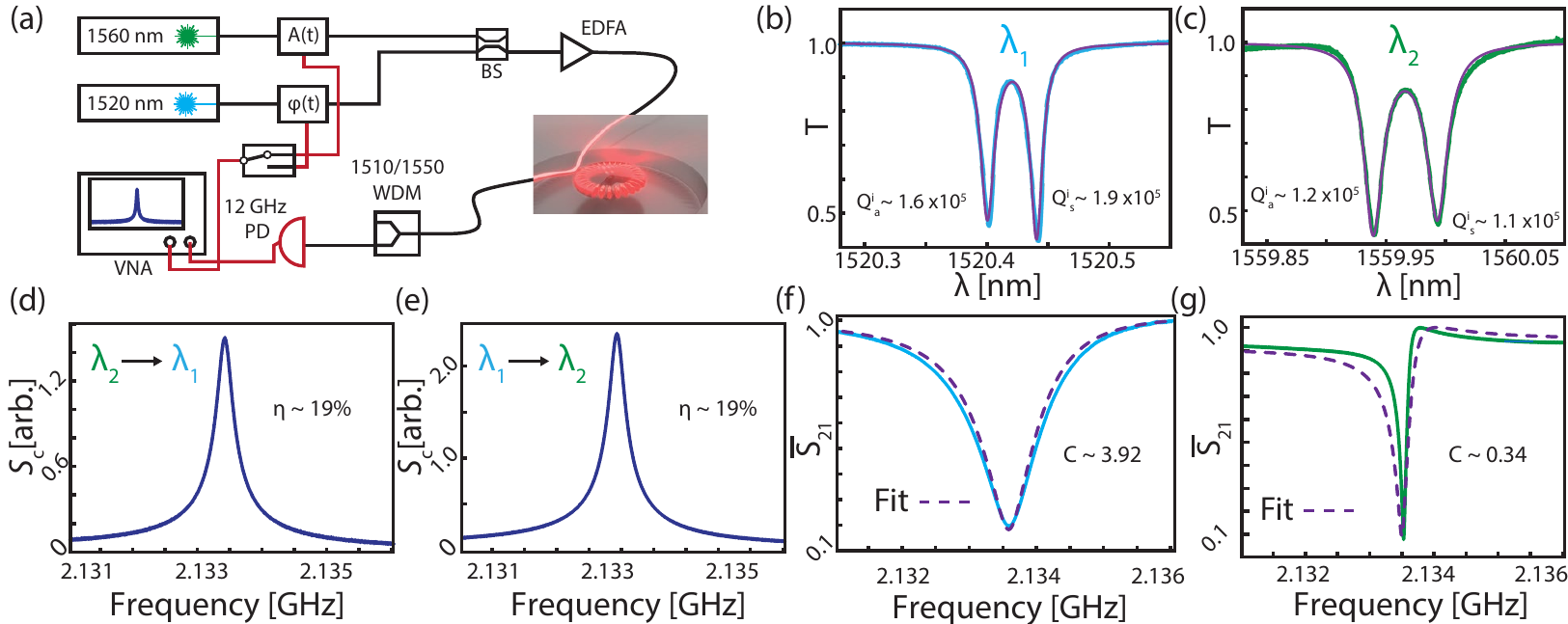}
 \caption{(a) Experimental setup used for frequency up-- and down--conversion. Phase and amplitude EOM's driven by the vector network analyzer (VNA) are used to generate the probe fields from the control fields where an RF switch controls which laser to modulate. A 50\%/50\% waveguide coupler combines the input fields which are coupled to the microdisk via a dimpled tapered fiber. A 1510/1550 WDM is used to filter the output of the cavity and the photodetected signal is analyzed by the VNA. (b,c) Optical whispering gallery mode resonances used in frequency conversion process. Intrinsic optical quality factors for the symmetric and anti--symmetric doublet modes labelled. (d,e) Beat note between converted photons and control field of the same color measured on the VNA for frequency up-- and down-- conversion respectively. (f,g) OMIT spectra for the $\lambda_1$ and $\lambda_2$ optical modes, respectively, where the cooperativity is extracted from fitting the OMIT lineshape.}
 \label{fig:Conversion}
\end{figure*}

\noindent In general the above may be written as

\begin{equation}
\left|t_\text{out}\right|^2 = |t^0|^2\alpha^2 + \alpha^2\beta|A|\cos(\omega t - \mathrm{arg}\{A\})
\end{equation}

\noindent and

\begin{equation}
\left|r_\text{out}\right|^2 = |r^0|^2\alpha^2 + \alpha^2\beta|B|\cos(\omega t - \mathrm{arg}\{B\}),
\end{equation}

\noindent where

\begin{equation}\label{eqn:A}
A = (t^{0*}t^+e^{-i\phi}+t^0t^{-*}e^{i\phi})
\end{equation}

\noindent and

\begin{equation}\label{eqn:reflectionFit}
B = (r^{0*}r^+e^{-i\phi}+r^0r^{-*}e^{i\phi}).
\end{equation}

The measured vector network analyzer (VNA) signal for the phase and amplitude quadrature are modeled by $\mathrm{arg}\{A\}$ ($\mathrm{arg}\{B\}$) and $|A|$ ($|B|$), respectively, for transmission (reflection) by the cavity. Additionally, by selecting the proper value of $\phi$ the response for a phase EOM, amplitude EOM, or mixture of both may be modelled. The bare cavity transmission amplitude $t^0(\Delta_\text{oc})$ is given by:

\begin{equation}\label{eqn:bare}
t^0(\Delta_\text{oc}) = 1 - \frac{\kappa_\text{e}/2}{i(\Delta_\text{oc})+\kappa/2}
\end{equation}

\noindent and for a red--detuned control field, the sideband transmission amplitudes as a function of $\Delta_\text{pc}$ are given by:

\begin{equation}\label{eqn:sidebands}
t^\pm(\pm\Delta_\text{pc}) = 1 - \frac{\kappa_\text{e}/2}{i(\Delta_\text{oc}-(\pm\Delta_\text{pc}))+\kappa/2+\frac{Ng_0^2}{i(\omega_\text{m}-(\pm\Delta_\text{pc}))+\Gamma_\text{m}/2}}.
\end{equation}

\noindent In the main text the reflected amplitude measured on the VNA is fit to $|B|$ by taking $r = 1 - t$, and substituting Eqns.\ \ref{eqn:bare} and \ref{eqn:sidebands} into Eqn.\ \ref{eqn:reflectionFit}, and choosing the appropriate value of $\phi$ depending on the EOM used. In order to determine $\phi$ for the amplitude EOM used in this work the phase difference ($2\theta$) between the sidebands was determined from the chirp parameter, $\alpha_\text{chirp}$, as:

\begin{equation}
\mathrm{tan}\theta \approx \alpha_\text{chirp} \mathrm{cot}\left(\frac{\varphi}{2}\right)
\end{equation}

\noindent assuming a small--modulation amplitude ($\beta \ll 1$), where $\varphi$ is the constant phase delay between the two interferometer arms, and $\varphi = \pi/2$ when biased at quadrature \cite{supp:ref:yan2003mot}. Through fitting the OMIT spectra obtained with the amplitude EOM in both the main text and in the following section, $\alpha_\text{chirp} \sim 0.70$ was found to provide the best fit; this value is within the bounds provided by the manufacturer.

For reference, the behavior of the model described above is shown in Fig.\ \ref{fig:EOM} for various $\phi$ and degree to which the system is sideband resolved. In particular it can be seen that for a non--pure amplitude modulator the observed VNA transmission spectra, $\bar{A} = |A|/\mathrm{max}(|A|$) can a resemble that of the reflection spectra,  $\bar{B} = |B|/\mathrm{max}(|B|$).

\section{Wavelength up-- and down--conversion}

For completeness, a demonstration of wavelength up-- and down-- conversion in the same device is presented here with an alternative microdisk on the same single--crystal diamond substrate as presented in the main text. This device exhibited two high-$Q$ optical modes that were within the operating range of a 1510/1550 nm wavelength division multiplexer (Montclair MFT-MC-51) which allowed the filtering of each mode from the fiber taper transmission, which we were unable to do with the device studied in the main text. Contrary  to the measurement presented in the main text this device was measured in transmission instead of reflection, as outlined in Fig.\ \ref{fig:Conversion}(a). Here two optical modes at $\lambda_1 = 1520$ nm and $\lambda_2 = 1560$ nm, as shown in  Fig.\ \ref{fig:Conversion}(b,c), were coupled to the fundamental radial breathing mode (RBM) at $\omega_\text{m}/2\pi = 2.135$ GHz, with $Q_\text{m}\sim 7,500$, for a similarly sized microdisk as studied in the main text. The measurement of both the up-- and down--converted signal with the vector network analyzer is shown in Fig.\ \ref{fig:Conversion}(d,e).

During this measurement symmetric or red mode of the $\lambda_1$ doublet was used such that the blue sideband of the EOM was off-resonance. However, the anti--symmetric or blue mode of the $\lambda_2$ doublet was used in the conversion process due it's the higher optical--$Q$. When setting up the experiment, balancing $C$ for each mode was performed by maximizing the contrast of each OMIT window, while attempting to reach $C_1 = C_2 > 1$. However, the relatively small splitting of this doublet led to a modification to the OMIT window spectral profile, due to participation of the red mode of the doublet, resulting in lower $C$ than expected based on the depth of the OMIT feature alone, and an overall low conversion efficiency due to the mismatch of $C_1$ and $C_2$. To account for this contribution the transmission amplitude for the blue sideband, $t^-(-\Delta_\text{pc})$, for an optical doublet was included as

 \begin{equation}
t^-(-\Delta_\text{pc}) = 1-\sqrt{\kappa_\text{e}/2}(a_\text{s}(-\Delta_\text{pc})+a_\text{a}(-\Delta_\text{pc})),
\end{equation}

\noindent where

\begin{equation}
a_\text{s,a}(-\Delta_\text{pc}) = \frac{-\sqrt{\kappa_\text{e}/2}}{-\kappa_\text{s,a}/2+i((\Delta_\text{oc}-(-\Delta_\text{pc}))\pm\kappa_\text{bs}/2)}
\end{equation}

\noindent are the symmetric ($a_\text{s}$) and anti-symmetric ($a_\text{a}$) combinations of the degenerate clockwise and counter--clockwise propagating travelling wave modes of the microdisk \cite{supp:ref:borselli2005brs}. Here $\kappa_\text{bs}$ is the backscattering rate between each mode, and $\kappa_\text{e}$ is taken to be equal for each doublet.

The OMIT spectrum obtained with the phase EOM shown in Fig.\ \ref{fig:Conversion}(f) was fit by including both side bands ($t^{\pm}(\pm\Delta_\text{pc}$) where a small phase delay ($< 5^{\circ}$) in sidebands was included to obtain good agreement with the spectra. In order to fit the OMIT spectrum obtained with the amplitude EOM shown in Fig.\ \ref{fig:Conversion}(g) both the inclusion of the chirp induced sideband phase difference $(2\theta\sim 70^{\circ})$ and the doublet transmission profile for ($t^{-}(-\Delta_\text{pc}$) were included. While the quality of this fit is inferior compared to the other OMIT fits, the inclusion of the doublet transmission profile reproduces the sharp, large contrast OMIT feature that in reality has $C<1$. This results in $C_1 \sim 3.92$, $C_2 \sim 0.34$, and $\eta_\text{int}\sim 15\%$, for $\eta_{1,2} \sim 19\%$,  resulting in $\eta_\text{ext} = 0.55\%$ for the device studied here. The fiber taper input power, $P_\text{in}$, was $\sim$ 17 mW and $\sim$ 4.2 mW, respectively, corresponding to $N_1\sim 7.0\times10^5$ and $N_2\sim 9.0\times10^4$ for $\lambda_1$ and $\lambda_2$.  Through the use of an additional EDFA it should be possible to reach $C_1 = C_2 > 1$, through independent control of $N$ for each mode; assuming that both modes could reach $C \sim 2$ would give $\eta_\text{int} \sim 64\%$.

\section{Optomechanical coupling calibration}

\begin{figure}[ht]
  \centering
  \includegraphics[width=\linewidth]{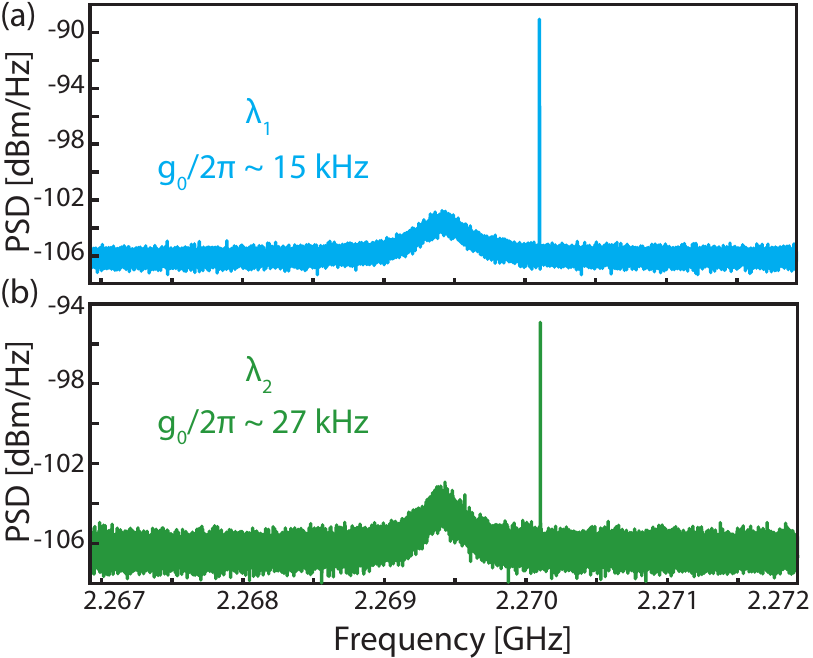}
 \caption{Phase tone calibration measurements for $\lambda_1$ (a), and $\lambda_2$(b) modes coupled to the fundamental RBM. Note that when computing the areas the PSD in units of mW/Hz was used.}
 \label{fig:Calibration}
\end{figure}

To estimate the vacuum optomechanical coupling rate of the $\lambda_1$ and $\lambda_2$ optical modes studied in the main text to the fundamental RBM the frequency noise method developed by Gorodetksy et al., which is now widely used in cavity optomechanics \cite{supp:ref:gorodetksy2010dvo,supp:ref:balram2016ccb,ref:schneider2016soc} was utilized. Here we compare the power spectral densities resulting from the thermomechanical cavity frequency fluctuations and a calibration tone generated with a phase modulator (EOSpace). By placing the frequency of the calibration tone, $\omega_\text{cal}$, close to the mechanical frequency, $\omega_\text{m}$, the transduction coefficients can be approximated as being equal, and $g_0$ can be calculated as

\begin{equation}
g_0 = \frac{\beta\omega_\text{cal}}{2}\sqrt{\frac{1}{n_\text{th}}\frac{S_\text{mech}(\omega_\text{m})}{S_\text{cal}(\omega_\text{cal})}}.
\end{equation}

\noindent Here $\beta = (V_\text{p}/V_\pi)\pi$ is the modulation index of the phase modulator, where $V_\text{p}$ is the amplitude of the driving RF signal, and $V_\pi$ is the modulators half wave voltage; $n_\text{th} = k_\text{B}T/\hbar\omega_\text{m}$ is the thermal occupation of the mechanical mode, and $S_\text{cal}(\omega_\text{cal}),S_\text{mech}(\omega_\text{mech})$ are the integrated powers in the calibration tone and mechanical mode response, obtained from the power spectral density measured on the real time spectrum analyzer in Fig.\ \ref{fig:Calibration}.

\section{Conversion gain estimation}

\begin{figure*}[ht]
  \centering
  \includegraphics[width=\linewidth]{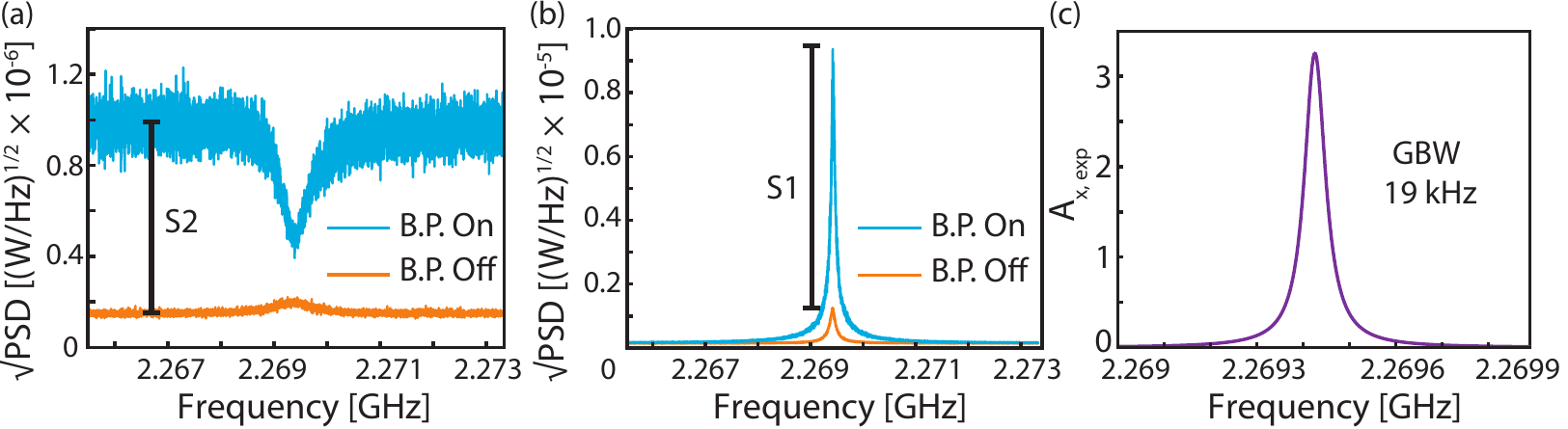}
 \caption{Estimation of amplified wavelength conversion gain, $A_\text{x, exp}$ utilizing the broadband probe (B.P.) measurement technique. (a) Measurement of cavity reflection and OMIT when the B.P. is on and transduction of the mechanical motion when the B.P. is off, where the $\lambda_1$ mode was placed off-resonance for this measurement. We define $S_2$ as the difference between the maximum of the cavity reflection profile (B.P. on) and the noise floor (B.P. off) which is measured slightly off--resonance to avoid the OMIT feature. (b) Amplified conversion measurement where we define  $S_2$ as the difference between the B.P. on and off measurement to isolate the wavelength converted signal from the thermal component (B.P. off). (c) Conversion gain calculated as a function of frequency where a maximum gain of $A_\text{x, exp} = 3.2$ is found, resulting in a GBW = 19 kHz.}
 \label{fig:Gain}
\end{figure*}

Here we present an experimental estimation of the amplified wavelength conversion gain, $A_\text{x, exp}$ using the broadband probe and RSA measurement technique discussed in the main text. In addition to the data shown in Fig. 4(c) of the main text, a measurement of the cavity reflection and OMIT for the $\lambda_2$ mode is utilized, as shown in Fig.\ \ref{fig:Gain}(a).  We define the estimated amplification conversion gain, $A_\text{x,exp}$, as the ratio of output converted signal power  $P_\text{out} = \alpha^2_{\lambda_{1,\text{conv}}}$ to the input signal power to be converted, $P_\text{in} = \alpha^2_{{\lambda_2}}\beta^2/4$ where $\alpha_{{\lambda_2}}$ is the control field amplitude for the $\lambda_2$ mode, and $\beta$ is the modulation index of the phase modulator driven by the arbitrary waveform generator. Following the methodology described in Section 1 the signal, $S_2$, or optical power, reflected by the cavity for the OMIT measurement of the $\lambda_2$ mode is

\begin{equation}
S_2 = \alpha^2_{{\lambda_2}}\beta|B|,
\end{equation}

\noindent where $B$ is defined in Eqn. \ref{eqn:reflectionFit}, and $\phi = \pi/2$. We then assume that the off resonant sideband reflection coefficient passes unaffected by the cavity ($r^-=0$) to write

\begin{equation}
S_2 = \alpha^2_{{\lambda_2}}\beta|r_{\lambda_2}^+||r_{\lambda_2}^0|,
\end{equation}

\noindent where the ${\lambda_2}$ subscripts have been added to the reflection coefficients to avoid confusion with the $\lambda_1$ mode. For the converted signal, $S_1$, as measured via the $\lambda_1$ mode, we take:

\begin{align}
r^+_\text{out} &= 0 \\ \nonumber
r^0_\text{out} &= r^0 \alpha_{\lambda_{1}} \\ \nonumber
r^-_\text{out} &= \alpha_{\lambda_{1,\text{conv}}}e^{-i(-\omega_\text{m}+\omega)t}
\end{align}

\noindent where $\alpha_{\lambda_{1}}$ is the control field amplitude for the $\lambda_1$ mode, which results in

\begin{equation}
S_1 = 2|r_{\lambda_1}^0|\alpha_{\lambda_{1}}\alpha_{\lambda_{1,\text{conv}}},
\end{equation}

\noindent which allows us to write the gain as

\begin{equation}\label{eqn:Gain}
A_\text{x,exp} = \frac{P_\text{out}}{P_\text{in}} = \frac{\alpha^2_{\lambda_{2}}|r_{\lambda_2}^0|^2}{{\alpha^2_{\lambda_{1}}|r_{\lambda_1}^0|^2}}\left(\frac{S_1}{S_2}\right)^2|r_{\lambda_2}^+|^2.
\end{equation}

\noindent We can then evaluate Eqn.\ \ref{eqn:Gain} based on the power we measure at the output of the cavity, the signals we measure on the RSA which are proportional to $S_1^2$ and $S_2^2$, and the reflection coefficient for the upper sideband input to the $\lambda_2$ mode, $r_{\lambda_2}^+$. To evaluate Eqn.\ \ref{eqn:Gain} we start by taking advantage of the fact that $\kappa \gg \gamma_\text{m}$ such that $r_{\lambda_2}^+$ is flat in the narrow spectral domain about the OMIT feature and set $r_{\lambda_2}^+ = \kappa_\text{e}/\kappa$, where we are approximating the on--resonance cavity reflection by measuring $S_2$ as shown in Fig.\ \ref{fig:Gain}(a). As the probe is derived from the $\lambda_2$ control field we are forced to measure the cavity reflection slightly detuned to avoid the OMIT feature in Fig.\ \ref{fig:Gain}(a);  this measurement is valid up to a constant which characterizes the electronic response of the photodetector. By assuming that the response of the photodetector is similar at both $\lambda_1$ and $\lambda_2$ this will be normalized out when computing $A_\text{x,exp}$. We define $S_1$ as the difference in the measured amplified signal with the broadband probe (B.P.) on and off as shown in Fig.\ \ref{fig:Gain}(b), which again ignores the photodetector response. The optical power measured at the output of the cavity for each mode is $\alpha^2_{\lambda_{1}}|r_{\lambda_1}^0|^2 = 16.0$ mW and  $\alpha^2_{\lambda_{2}}|r_{\lambda_2}^0|^2 = 3.8$ mW. The signals in Fig.\ \ref{fig:Gain}(b) were each fit to a Lorentzian which were then used to calculate $A_\text{x, exp}$ as a function of frequency, as shown in  Fig.\ \ref{fig:Gain}(c). This gives a maximum gain of $A_\text{x,exp} = 3.2$, and bandwidth of 5.8 kHz, resulting in a gain bandwidth product, GBW = 19 kHz.

\end{document}